\documentstyle[12pt]{article}
\begin{document}
\begin{titlepage}
\begin{center}
{\LARGE\bf The Physics Inside\\
Topological Quantum Field Theories\footnote{
This work is supported in
part by funds provided by the
U. S. Department of Energy (D.O.E.) under cooperative agreement
\#DE-FC02-94ER40818.
}}
\end{center}
\vskip 0.3truecm
\begin{center}
{
{\Large R}{OGER} {\Large B}{ROOKS}\footnote{On leave from the Physics
Department, Massachusetts Institute of Technology, Cambridge, MA 02139,
U.S.A.}
}
\end{center}
\begin{center}
{\it {Department of Physics,}}\\
{\it {Stanford University}}\\
{\it {Stanford, California 94305, U.S.A.}}
\end{center}
\vskip 0.5truein
\begin{abstract}
We show that the equations of motion defined over a specific field space
are realizable as operator conditions in the physical sector of a generalized
{}Floer theory defined over that field space.  The ghosts associated with such
a construction are found not to be dynamical. This construction is applied to
gravity on a four dimensional manifold, $M$; whereupon, we obtain Einstein's
equations via surgery, along $M$, in a five-dimensional topological quantum
field theory.
\end{abstract}
\vskip 0.5truein
\leftline{SU-ITP-95-28 \hfill November 1995}
\leftline{MIT-CTP-2491}
\smallskip
\leftline{hep-th/9512003}
\end{titlepage}


Within the last year, a number of authors have espoused the view that at least
a subset of the physics of four-dimensional quantum gravity may be
realized in a topological quantum field theory (TQFT)
\cite{{BroLif},{Crane},{Smol}}. However, only partial success in
establishing this statement has been obtained thus far.  Unlike previous
work, we will focus our attention on a five-dimensional TQFT in order to
realize four-dimensional gravity.  The clue which we will exploit is that
for a TQFT on a manifold with boundary, the
``classical'' fields are dynamical on the boundary whereas the ghosts are
not. Furthermore and of equal importance, is the absence of dynamical
degrees of freedom in the bulk (compliment of the boundary) of the
five-dimensional manifold.  With Atiyah's axioms and Floer cohomology
theory in mind, we are then  lead to the examination of the quantum
mechanics of a five-dimensional TQFT.

Our note is organized as follows.  First, we will derive some results
which follow as essentially straightforward exercises/extrapolations from
the existing literature \cite{TQFTrev}.  Secondly, we realize the
connection of these results with Einstein's equations. Finally, we discuss
the covariant version of the five-dimensional TQFT. Four dimensional
space-time will be denoted as $M$ while the five-dimensional manifold will
be written as $X$.

To start, we assume
that we are given the equations of motion of a physical field theory (or
any field theory for that matter) for a set of fields, $\Phi^I$, and that
these are obtained by extremizing the action $S_{ft}[\Phi^I]$.  We now show
that:

\vskip 6pt
{\sl
{}Formally, a TQFT on a $(d+1)$-dimensional manifold, $X$  exists such that
the equations of motion for the $\Phi^I$ on a $d$-dimensional manifold, $M$,
arise as operator conditions in the physical sector of Hilbert space.
{}Furthermore, in this Hilbert subspace, the only dynamical degrees of freedom
on $M$ are those of the $\Phi^I$.
}
\vskip 6pt

As was the case with Floer's treatment of holomorphic maps \cite{Floer1}
and flat connections \cite{Floer2}, we identify  $S_{ft}$ as the
Morse functional on the space, ${\cal U}$, of these fields.  From this we
introduce a second set of fields, $\psi^I$, which carry the same spin,
but opposite statistics to the $\Phi^I$ and define the exterior derivative,
$Q$, and its adjoint, $Q^*$, on ${\cal U}$:
\begin{eqnarray}
Q&~=~ \langle (\frac{\delta~}{\delta \Phi^I} ~+~ \frac{\delta
S_{ft}[\Phi]}{\delta \Phi^I}), \psi^I\rangle\ \ ,\nonumber \\
Q^*&~=~ -\langle \chi^I, (\frac{\delta~}{\delta \Phi^I} ~-~
\frac{\delta
S_{ft}[\Phi]}{\delta \Phi^I})\rangle\ \ ,
\label{Qft}
\end{eqnarray}
where the bracket represents the inner products on $M$ and ${\cal U}$, while
$\chi^I$ is the canonical momenta (dual) of $\psi$.  Lastly, define the
Hamiltonian of the TQFT to be \cite{WitTYM}
\begin{equation}
H_{TQFT}~=~ {1\over 2}  \{Q^*,Q\}\ \ .
\label{HTQFT}
\end{equation}
With these definitions, we can now proceed to establish our statement.

A choice of hermitian conjugation exists for which $H_{TQFT}$
is hermitian.
By standard arguments it follows that the physical Hilbert space of the
TQFT is ${\cal H}_{TQFT} \in {\rm Ker}(Q)/{\rm Im}(Q)$ and are ground states of
the
Hamiltonian.  Let $|{\rm phys}_i\rangle$ denote a basis element for this
space so that $\langle {\rm phys}_i|{\rm phys}_j\rangle=\delta_{ij}$.
It then follows that for any operator ${\cal O}$,
all physical state matrix elements of the commutator $[H_{TQFT},{\cal O}]$,
vanish.
In particular, the matrix elements of the momenta conjugate to the
$\Phi^I$ fields vanish:
\begin{equation}
\langle \hbox{phys}_i|\pi_I|\hbox{phys}_j\rangle ~=~ 0\ \ ,
\label{HamO}
\end{equation}

The vanishing of the momenta is consistent with the absence of dynamical
degrees of freedom in the TQFT.  Furthermore, this means that as a
composite operator,
$\frac{\delta S_{ft}}{\delta\Phi^I}$ is represented by $0$  on the physical
subspace since its matrix elements vanish:
\begin{equation}
\langle {\rm phys}_i| \frac{\delta S_{ft}[\Phi]}{\delta \Phi^I}|
{\rm phys}_j\rangle
{}~=~ 0
\qquad
\Longrightarrow
\qquad
\frac{\delta S_{ft}}{\delta\Phi^I}~=~ 0\ \ .
\label{EQMvev}
\end{equation}
{}Finally, given that $\psi^I$ is $Q$-exact, its matrix elements between any
physical states vanish:
\begin{equation}
\langle {\rm phys}_i| \psi^I |
{\rm phys}_j\rangle ~=~ 0 \ \ .
\label{psiev}
\end{equation}
We have established the statement.

The exterior derivative defined in Eqn. (\ref{Qft}) actually follows as
the BRST charge from the action
\begin{equation}
I_0[\Phi]~=~ \int_X dL\ \ ,
\end{equation}
where $L$ is what we might normally write down (but lifted to X) as a
Lagrangian density for $\Phi$: $S_{ft}\equiv \int_M L$.  In this regard,
the four-dimensional
manifold $M$ is the boundary of some $X$ if its signature vanishes:
$\sigma(M)=0$ \cite{Rohlin}. If we take $X$ to be
the semi-infinite cylinder with time normal to the boundary $M$,
we then find that the momentum, $\pi_I$, conjugate to $\Phi^I$ is given by the
constraint
\begin{equation}
\pi_I ~-~ \frac{\delta }{\delta \Phi^I}\int_{M} L ~\approx~0\ \ .
\label{CONSTR}
\end{equation}
Then $Q$ follows by BFV-quantization with this constraint.  Thus our
definitions (\ref{Qft}) are not absolutely necessary. However, we prefer
to start with $Q$ rather than $I_0$ as we have seen that all we need
is the canonical theory to establish our results.  The
construction of Hamiltonia for TQFT's wherein the constraints in the BFV
charges were taken to be the ADM-constraints of quantum gravity theories,
has been discussed \cite{Bro}.

Let us now specialize to gravity.  The functional, $S_{ft}$,
in this case is the
Einstein-Cartan action which we treat as a gauge theory
with gauge group $SO(3,1)$ (indices $a,b,\ldots$). The
fields of this theory are one-forms $e^a$ and
$\omega^{ab}$ on $M$;  these are the vierbien and Lorentz spin-connection
of the four-dimensional theory.  The space of these fields will be denoted
as ${\cal U}$.
We lift the fields to $X$ where they continue to carry the same gauge
group
representations but vastly different physical interpretations; we will
adorn the lifted fields with $\hat{}$'s: $({\hat e}^a,{\hat\omega}^{ab})$.  The
vierbien becomes ${\hat e}^a$ which is a peculiar matter field on $X$ that,
as we will see below, is
coupled to the curvature of a $SO(3,1)$ connection,
${\hat\omega}^{ab}$. The space of these lifted fields will be denoted as
$\hat{\cal U}$.
In addition to these fields, we also
introduce a pair of Grassmann-odd one-forms $\zeta^a$ and
$\xi^{ab}$ along with their lifts $\hat\zeta^a$ and $\hat\xi^{ab}$. These
are in $T^*{\cal U}$ and $T^*{\hat{\cal U}}$, respectively.

The exterior derivative on the space of
fields which we will use is
\begin{eqnarray}
Q&~=~& e^{-\,S_{EC}}~\int_M Tr[
\zeta\wedge \,{^*{\hskip -1.5pt}} \frac{\delta}{\delta \omega} ~+~
\xi\wedge \,{^*{\hskip -1.5pt}} \frac{\delta}{\delta e}
]~ e^{\,S_{EC}}\ \ ,\nonumber \\
S_{EC}&~=~& \int_M e^a\wedge e^b\wedge F^{cd}(\omega)\epsilon_{abcd}\ \ ,
\label{QECFLOER}
\end{eqnarray}
where $S_{EC}$ is
the Einstein-Cartan action
so that a critical point is specified by Einstein's equations.  Then a
wavefunctional
which  satisfies the functional equations
\begin{equation}
\frac{\delta}{\delta \omega^{ab}} ~-~ \, {^*{\hskip -1.5pt}} \, (e^c\wedge
{\cal D} e^d)
\epsilon_{abcd}
{}~=~ 0\qquad \hbox{and}\qquad
\frac{\delta}{\delta e^a} ~+~ \,  {^*{\hskip -1.5pt}}\, (e^b\wedge
{}F^{cd})\epsilon_{abcd}~=~ 0\ \ ,
\label{ECONSTR4D}
\end{equation}
is an element of $\ker{Q}$.  By construction, the $\bf 1$ in the
cohomology of $Q$ is
\begin{equation}
\Psi_{1}[e,\omega]~=~ e^{-S_{EC}[e,\omega]}\ \ .
\label{Psi4D}
\end{equation}
The Einstein-Cartan (hence Einstein's) equations of motion on $M$ follow
as per our discussion above:
\begin{equation}
e^a\wedge {\cal D} e^b\epsilon_{abcd} ~=~ 0
\qquad\hbox{and}\qquad
e^a\wedge F^{bc}\epsilon_{abcd} ~=~ 0\ \ .
\label{EEQM4D}
\end{equation}
Although a metric on $X$ was needed in the construction of $Q$ it drops
out when Eqn. (\ref{EQMvev}) is implemented.  Consequently, in focusing
only on the
equations in (\ref{EEQM4D}) we are free to identify $e^a$ with the vierbien
just as we do in the usual discussion of the Einstein-Cartan action.  Of
course, this is a residue of the topological nature of the theory on $X$
which we now turn to.

We can
deduce Eqns. (\ref{ECONSTR4D}) and (\ref{EEQM4D}) by starting with the
classical five-dimensional action:
\begin{equation}
I_0[{\hat e},{\hat \omega}]~=~
\int_{X} {\hat e}^a\wedge {\hat T}^b\wedge {\hat
{}F}^{cd}\epsilon_{abcd}\ \ ,
\qquad {\hat T}~\equiv~ {\hat {\cal D}}{\hat  e}\ \ .
\label{EEC}
\end{equation}
Writing this action on $X=M\times R$, we see that
Eqn. (\ref{ECONSTR4D}) follows from the computation of the canonical
momenta of the fields $\omega$ and $e$.
The covariant quantum theory on arbitrary $X$, follows by imposing the forty
gauge fixing conditions
\begin{equation}
{\hat T}_d~=~ {^*{\hskip -1.5pt}}{\hat  F}^{ab}\wedge {\hat e}^c
\epsilon_{abcd}\ \ ,
\label{EGF}
\end{equation}
on the fifty fields in $\hat{\cal U}$. We have introduced the symbol ${\hat T}$
as the
restriction, $T$, of this $2$-form to four dimensions is in fact the torsion of
$M$.  Notice that the four-dimensional Einstein tensor and torsion are
not related by this equation.  Rather, equation (\ref{EGF}) equates the
Einstein tensor with the derivative of the would be
vierbien with respect to the extra fifth coordinate.
Much as in the previous section, we can write down the purely bosonic part
of the gauge fixed TQFT action as
\begin{equation}
I_{bos}~=~ \int_{X} Tr[{1\over 2} (T ~-~ {^*{\hskip -1.5pt}} F\wedge e)^2]\
\ .
\end{equation}
This functional plus its ghost completion and $I_0$ is the
five-dimensional quantum action for four dimensional classical gravity.

We are assured that the TQFT whose moduli space is defined by
Eqn. (\ref{EGF}) is non-trivial by virtue of the fact that we can
identify some of its subspaces.  One subspace which is immediate is given
by $X=M\times S^1$ and using the five-dimensional diffeomorphism and
gauge invariance of Eqn. (\ref{EGF}) to set
${\hat e}^a$ and ${\hat \omega}^{ab}$
in the $S^1$
direction to zero.  Then we see that $M$ is generically an Einstein space
as the $S^1$ harmonics of the $M$ frame $e^a$ result in a (quantized)
cosmological constant.

In summary, we have discovered that the equations of motion of a
$d$-dimensional field theory follow from a $(d+1)$-dimensional
TQFT via surgery.  Importantly, the ghosts of the TQFT vanish on
the $d$-dimensional manifold.  The explicit example of four dimensional
gravity was given. Another, and perhaps more immediate, example of our
construction is the realization of the
three dimensional Chern-Simons equations of motion,
from the four dimensional Pontryagin density
via surgery. Recall that this is the
root of the flows of flat connections as described by Floer
\cite{Floer2}.

We infer from these results that classical physics appears in the midst of
pure topology.  Alternatively,
we can arrange for the equations of motion of a field theory on a manifold $M$
to arise as boundary conditions for a TQFT on a manifold $X$ with
boundary $\partial X=M$.  Thus the classical physics on a manifold
follow from a functor from the category of those manifolds into the category of
physical Hilbert spaces of the TQFT. We can take this further by using results
from the axiomatization of quantum field theories \cite{Segal} to realize such
processes as spacetime (as opposed to spatial) topology change by taking
$\partial X =M\cup (-M')$.  As the $(d+1)$-dimensional theory whose
$d$-dimensional ``spaces"
(which we in turn realize as spacetimes) are undergoing
topology change is topological, we are free to relax the isochronous condition
\cite{Geroch} as causality is not an issue here. Degenerate metrics on $X$
are allowed.

\vskip 0.5truein
\leftline{\bf Acknowledgements}
The authors thanks R. Kallosh and the Physics Department at Stanford
University for its hospitality.



\begin{thebibliography}{99}

\bibitem{BroLif}{R. Brooks and G. Lifschytz, {\it Nucl. Phys.} {\bf B438}
(1995) 211.}

\bibitem{Crane}{L. Crane, {\it J. Math. Phys.} {\bf 36} (1995) 6180.}

\bibitem{Smol}{L. Smolin, {\it J. Math. Phys.} {\bf 36} (1995) 6417.}

\bibitem{TQFTrev}{D. Birmingham, M. Blau, M. Rakowski and G. Thompson,
{\it Phys. Reps.} {\bf 209} (1991) 129 and references therein.}

\bibitem{Floer1}{A. Floer, {\it Bull. Am. Math. Soc.}{\bf 16} (1987) 279.}

\bibitem{Floer2}{A. Floer,{\it Commun. Math. Phys.} {\bf 118} (1988) 215.}

\bibitem{WitTYM}{E. Witten, {\it Commun. Math. Phys.} {\bf 117} (1988) 353.}

\bibitem{Rohlin}{For a contemporary treatment of this theorem (originally
due to V. A. Rohlin) see P. Melvin, {\it Contemp. Math.} {\bf 35} (1984) 399.}

\bibitem{Bro}{R. Brooks, {\it Mod. Phys. Letts.} {\bf 8} (1993) 2277.}

\bibitem{Segal}{G. Segal, {\sl Two-dimensional conformal field theories
and modular functors}, in the proceedings of the IX$^{\rm th}$
International Congress on Mathematical Physics, July 17 - 27, 1988,
Swansea, Wales; B. Simon, A. Truman and I. M. Davies, eds.; Adam Hilger,
Bristol, 1989.}

\bibitem{Geroch}{R. Geroch, {\it J. Math. Phys.} {\bf 8} (1967) 782}

\end{thebibliography}
\end{document}